\documentclass[prc,aps,amsfonts,twocolumn,showpacs,floatfix,dvips]{revtex4}
\usepackage{graphics,epsfig,amssymb}
\newcommand{\bra}[1]{{\langle{#1}}}
\newcommand{\ket}[1]{{\vert{#1}\rangle}}
\newcommand{\be}{\begin{equation}}
\newcommand{\ee}{\end{equation}}
\newcommand{\bqr}{\begin{eqnarray}}
\newcommand{\eqr}{\end{eqnarray}}
\newcommand{\ba}{\begin{array}}
\newcommand{\ea}{\end{array}}
\newcommand{\bt}{\begin{tabular}}
\newcommand{\et}{\end{tabular}}
\newcommand{\bc}{\begin{center}}
\newcommand{\ec}{\end{center}}

\begin{document}
\preprint{APS/123-QED}

\title{\Large \bf Microscopic approach of
fission dynamics    \\
applied to fragment kinetic energy   \\
and mass distributions in $^{238}$U}

\author{H. Goutte},
\email{heloise.goutte@cea.fr}
\author{J.F. Berger}
\author{P. Casoli},
\author{D. Gogny}
\altaffiliation{Present address: L-414 Lawrence Livermore National Laboratory, Livermore, California 94551, USA}
\affiliation{CEA/DAM Ile de France
DPTA/Service de Physique Nucl\'eaire, \\
BP 12, 91680 Bruy\`eres-le-Ch\^atel, France}
\date{\today}

\begin{abstract}
The collective dynamics of low energy fission in $^{238}$U is described
within a time-dependent formalism based on the Gaussian Overlap
Approximation of the time-dependent Generator Coordinate Method.
The intrinsic deformed configurations of the nucleus are determined from the self-consistent
Hartree-Fock-Bogoliubov procedure employing the effective force D1S with constraints on the quadrupole and
octupole moments. Fragment kinetic energy and mass distributions are calculated and compared with experimental
evaluations. The effect of the collective dynamics along the fission paths and the influence of
initial conditions on these distributions are analyzed and discussed.
\end{abstract}


\maketitle

%%%%%%%%%%%%%%%%%%%%%%%%%%%%%%%%%%%%%%%%%%%%%%%%%%%%%%%%%%%%%%%%%%%%%

\section{\label{sec1}Introduction}

Interest in fission has recently increased since it is proposed to be used in new applications, such as accelerator
driven systems, new electro-nuclear cycles - thorium based fuel cycle -, and the next generation of exotic beam facilities.
For these applications, there is an important need for fission cross-sections in a large range of excitation energies,
and also for mass-charge fission fragment distributions.
For instance, precise knowledge of production rates of secondary long-lived fission residues and of neutron-rich isotopes
is crucial for designing and simulating these new facilities. It is worth pointing out that relevant measurements of mass
and charge distributions have been performed recently.
For instance the production of exotic nuclei has been measured from spallation reactions of 1~A GeV $^{238}$U projectiles
on an hydrogen target~\cite{Be03} and isotopic yields have been deduced for elements between $^{58}$Ni and $^{163}$Eu.
Furthermore, thanks to secondary beam facilities, fission properties of 70 short-lived radioactive nuclei can be found
in references~\cite{Sc00,Be02}. Such a systematic analysis of the fission properties covers a wide region of
the nuclide chart and the transition between single- and double-humped mass distributions has been observed with a
triple-humped structure for $^{227}$Th.
It is important to test the accuracy of the theoretical prediction using the data in order to gain confidence in its
predictions
when applied to widely extended domains such as fission of nuclei far from stability, and fission for a large range
of excitation energies. \\
From a theoretical point of view, the description of fission process stands at the crossroads of many subjects in the
forefront of research.
Both static and dynamical properties of the fissionning system are required, namely: nuclear configurations far from
equilibrium, the interplay of collective and intrinsic degrees of freedom, and the dynamics of large amplitude collective
motion. Theoretical works generally focus on the static part of the fission. For instance, many studies have been devoted to
multi-dimensional potential energy surfaces~\cite{Mo00,Pa88} from which fission barriers are extracted
and to nuclear configurations at scission and associated fragment distributions~\cite{Wi76,Br88}.
On the other hand there are very few dynamical studies
of fission, although dynamical effects are expected to play an essential role in particular in the descent from saddle
to scission.
Fragment mass distributions have recently been obtained by solving the classical three-dimensional Langevin
equations~\cite{Ka01}. The influence of the mass asymmetry degree of freedom on the variance of the mass distribution has
been highlighted. Two types of microscopic quantum dynamical calculations have also been performed in the past. First, in
1978 the time-dependent Hartree Fock method~\cite{Ne78} has been applied to fission. Second, time-dependent calculations
based on the Generator Coordinate method
using Hartree-Fock-Bogoliubov states have been performed, and the most probable fission configuration of $^{240}$Pu has been
analyzed~\cite{Be84}. The present study is an extension of this pioneering work. \\
In the present work we have chosen to derive the collective dynamics of
fission using a time-dependent formalism based on the Gaussian Overlap
Approximation of the time-dependent Generator Coordinate (GC) theory.
An alternate method could have been to first determine the
stationary solutions of the GC equations within the relevant domain of generator coordinates with appropriate boundary
conditions. The solutions of the time-dependent GC equations would then be expressed in a straightforward manner.
However, the precise form of the boundary conditions to be used is difficult to obtain when more than one generator
coordinate are employed. Applying a time-dependent method allows one to avoid this problem. The only input of the
calculation is the collective wave function chosen at t=0. Spurious reflections of the time-dependent collective
wave function on the edge of the finite domain are eliminated using a standard absorption technique as explained in
Section~\ref{sec3}.  \\
In this paper, we focus on low energy fission-fragment distributions of $^{238}$U, and also on several physical aspects
that can be clearly analyzed in this even-even fissioning system.
Let us recall that, at low energy, elongation and asymmetry degrees of freedom are among the most relevant ones and that the
adiabatic assumption is to a large extent justified~\cite{Wa91}.
As we explain below, time evolution in the fission channel is described in terms of a wave function of Hill-Wheeler type.
The latter is taken as a linear combination of Hartree-Fock-Bogoliubov (HFB) solutions characterized by the two
collective degrees of freedom just mentioned. It is worth pointing out that this work relies only on the D1S effective
interaction used at Bruy\`eres-le-Ch\^atel.
\\
The calculation proceeds in two steps: the potential energy surface and the collective inertia are determined from the
first well to scission and then, the dynamical treatment of fission is performed using an approximate Time
Dependent Generator Coordinate Method (TDGCM). Potential energy surfaces and associated collective inertia tensors
are calculated using the constrained Hartree-Fock-Bogoliubov approach with the D1S finite-range effective
force~\cite{DG80,Be91}. Fission wave functions at time zero are constructed from the quasi-stationary collective states
in the first well. Their time evolution is calculated numerically by discretizing on a mesh a time-dependent
Schr\"odinger-like equation.
Mass distributions are derived from the flux of the wave function through scission at given A$_H$/A$_L$ fragmentations.   \\
The present work is organized as follows. The HFB formalism and the TDGCM
method are presented in Section~\ref{sec2} and numerical procedures are detailed in section~\ref{sec3}.
Section~\ref{sec4} is devoted to the static results,
where the potential energy surface and pairing correlations are discussed. A first estimate of kinetic energy and fragment
mass distributions, obtained from a "static" calculation at scission, are discussed.
Mass distributions
obtained from the full time-dependent calculations are presented in Section~\ref{sec5}, and the crucial role
played by dynamical effects
is analyzed.
%%%%%%%%%%%%%%%%%%%%%%%%%%%%%%%%%%%%%%%%%%%%%%%%%%%%%%%%
\section{\label{sec2}Formalism}
In low energy fission the adiabatic hypothesis seems to be justified~\cite {Wa91} and, therefore, collective and intrinsic
degrees of freedom can be decoupled. Furthermore, we assume that the collective motion of the system can be described
in terms of a few collective variables characterizing the shape evolution of the nucleus. In a self-consistent
formalism these shapes can be generated by means of
external fields represented by the operators:
\be
\displaystyle \hat{Q}_{20}    =  \sqrt{\frac{16\pi}{5}}\sum_{i=1}^{A}r_i^2Y_{20}  =\sum_{i=1}^{A}(2z_i^2-x_i^2-y_i^2)  ,
\label{q23}
\ee
and
\be
\displaystyle \hat{Q}_{30}    =  \sqrt{\frac{4\pi}{7}}\sum_{i=1}^{A}r_i^3Y_{30}   =\sum_{i=1}^{A}(z_i^3-\frac{3}{2}
z_i(x_i^2+y_i^2)) .
\label{qq23}
\ee
These moments govern mass axial deformation and left-right asymmetry of the nucleus, respectively.
For well-separated fragments one can express the mean values of these operators in terms of $<Q_{20}^H>$ ($<Q_{30}^H>$)
and $<Q_{20}^L>$ ($<Q_{30}^L>$), the mean quadrupole (octupole) deformations of the heavy and light fragments, respectively,
$d_m$ the distance between their centers of mass, and~$\mu$~the~reduced~mass~:
\be
\displaystyle
\begin{array}  {lll}
 <Q_{20}>   & = & \displaystyle <Q_{20}^H> + <Q_{20}^L> + 2\mu d_m^2  \ ,\\
            &   &    \\
 <Q_{30}>   & = & <Q_{30}^H> + <Q_{30}^L>  \\
            &   & \displaystyle + \frac{3d_m}{A_H + A_L}(A_H <Q_{20}^L> - A_L <Q_{20}^H>) \\
            &   & \displaystyle +   2\mu d_m^3 \frac{A_H - A_L}{A_H + A_L}   \ ,
\end{array} \label{q23b}
\ee
with
\be
\mu=\frac{A_H  A_L}{A_H + A_L} .
\label{mu}
\ee
Relations~(\ref{q23b}) and~(\ref{mu}) have only been used in the present work to check the validity of the computer program
for configurations close to scission.  \\
The intrinsic axially-deformed states $\ket{\Phi(q_{20},q_{30})}$ of the fissile system are taken as the solutions
of the constrained
Hartree-Fock-Bogoliubov variational principle~\cite{Be91}:
\begin{equation}
\displaystyle \delta \bra{\Phi(q_{20},q_{30})}|\hat{H}-\lambda_N \hat{N} -\lambda_Z \hat{Z}-\sum_i \lambda_i \hat{Q}_i
\ket{\Phi(q_{20},q_{30})}=0   \ ,
\label{eq2}
\end{equation}
the Lagrange parameters $\lambda_N$, $\lambda_Z$, and $\lambda_i$ being deduced from:
$$\bra{\Phi(q_{20},q_{30})}|\hat{N} \ket{\Phi(q_{20},q_{30})}  =  N  ,$$
\be
\bra{\Phi(q_{20},q_{30})}|\hat{Z} \ket{\Phi(q_{20},q_{30})}  =  Z  ,
\label{eq3}
\ee
and
$$\bra{\Phi(q_{20},q_{30})}|\hat{Q}_i \ket{\Phi(q_{20},q_{30})}  =  q_i.$$
In Eq.~(\ref{eq2}), $\hat{Q}_i$ is the set of external field operators ($\hat{Q}_{20}$, $\hat{Q}_{30}$, $\hat{Q}_{10}$),
and $\hat {H}$ is the nuclear many-body effective Hamiltonian built with the finite-range effective
force D1S~\cite{Be91}. The additional constraint on the dipole mass operator is used in order to fix the position
of the center of mass of the whole system. This is accomplished by setting $< \hat{Q}_{10} >$~=~0, where:
\be
\displaystyle
\begin{array}  {llll}
\displaystyle \hat{Q}_{10}   & = & \displaystyle \sqrt{\frac{4\pi}{3}}\sum_{i=1}^{A}r_iY_{10} & \displaystyle
=\sum_{i=1}^{A}z_i \ .
\end{array} \label{q10}
\ee
The system of Eqs.~(\ref{eq2}) and~(\ref{eq3}) is solved numerically for each set of deformations by expanding the single
particle states onto an axial harmonic oscillator (HO) basis. For small elongation, 0 $< q_{20} \le$ 190 b, one-center bases
with N~=~14 major shells have been considered, whereas for well-elongated configurations $q_{20}  > $ 190 b, two-center bases
with N~=~11 for each displaced HO basis have been used.
Because calculations are performed in an even-even nucleus for which K~=~0 -- with K the projection of the spin onto the
symmetry
axis --, the HFB nuclear states are even under time-reversal symmetry $\hat{T}$. Furthermore, we restrict the Bogoliubov
space by imposing the self-consistent symmetry
$\hat{T}\hat{\Pi}_2$, where $\hat{\Pi}_2$ is the
reflection with respect to the xOz plane. Let us mention that the octupole operator breaks the parity symmetry. However,
since $\hat{P}\hat{Q}_{30}\hat{P}^{-1}$ = $-\hat{Q}_{30}$ and $\hat{P}\hat{H}\hat{P}$ = $\hat{H}$ with $\hat {P}$ the
parity operator, constrained HFB calculations can be restricted to positive values of $q_{30}$, and negative ones
are obtained from $\ket{\phi(q_{20},-q_{30})}=\hat{P}\ket{\phi(q_{20},q_{30})}$.

The nucleus time-dependent state is defined as a linear combination of the basis states $\ket{\phi(q_{20},q_{30})}$:
\begin{equation}
\displaystyle
\ket{\Psi (t)}  = \int \int dq_{20} \;  dq_{30} \; f(q_{20},q_{30},t) \; \ket{\phi(q_{20},q_{30})} \ ,
\label{eq1}
\end{equation}
where
$f(q_{20},q_{30},t)$ is a time-dependent weight function which is
obtained by applying the variational principle:
\begin{equation}
\displaystyle
\frac{\delta}{\delta f^*(q_{20},q_{30},t)}\int_{t_1}^{t_2} \bra{\Psi(t)}|\hat{H}-i\hbar \frac{\delta}{\delta t}
\ket{\Psi(t)}dt = 0  \ ,
\label{eqtime}
\end{equation}
where $\hat{H}$ is the same microscopic Hamiltonian as the one introduced in Eq.~(\ref{eq2}).
The result is the well-known Hill-Wheeler equation which reduces to a time-dependent Schr\"odinger equation when the
GCM problem is solved using the Gaussian Overlap Approximation (GOA)~\cite{Li99}:
\begin{equation}
\displaystyle
\hat{H}_{coll} \; g(q_{20},q_{30},t)= \; i\hbar \; \frac{\partial g(q_{20},q_{30},t)}{\partial t} \ .
\label{eq5}
\end{equation}
The collective wave functions $g(q_{20},q_{30},t)$ solutions of Eq.~(\ref{eq5}) are related to the weight functions
$f(q_{20},q_{30},t)$ through the following relation:
\be
\begin{array}{lll}
\displaystyle
g(q_{20},q_{30},t) & = & \int \int dq'_{20}dq'_{30}f(q'_{20},q'_{30},t)\\
                   &   & I^{\frac{1}{2}} (q_{20},q_{30},q'_{20},q'_{30})  \ ,
\end{array}
\label{timestate}
\ee
where $I^{\frac{1}{2}}(q_{20},q_{30},q'_{20},q'_{30})$ is the square root kernel of the overlap kernel:
$$\displaystyle I(q_{20},q_{30},q'_{20},q'_{30})=\bra{\phi(q_{20},q_{30})}\ket{\phi(q'_{20},q'_{30})}.$$

The exact form of the collective Hamiltonian $\hat{H}_{coll}$ deduced
from the GOA can be found in~\cite{RS,Ro88}. In the present derivation
of this Hamiltonian, the widths $G_{22}$, $G_{23}$ and $G_{33}$ of the
gaussian overlap between differently deformed constrained HFB states
have been assumed to be constant. Numerical calculation of these widths
shows that they vary very slowly in the whole $q_{20}$--$q_{30}$
domain considered here and that their variations can be
neglected. With this assumption, the two-dimensional collective
Hamiltonian reads:
\begin{equation}
\begin{array}{lll}
\displaystyle \hat{H}_{coll} & = &\displaystyle
-\frac{\hbar^2}{2}\sum_{i,j=2}^{3}\frac{\partial} {\partial
q_{i0}}B_{ij}(q_{20},q_{30}) \frac{\partial}{\partial
q_{j0}}+V(q_{20},q_{30})\\
& &\displaystyle -\sum_{i,j=2}^{3}\Delta V_{ij}(q_{20},q_{30})  \ ,
\end{array}\label{eq6}
\end{equation}
where $V(q_{20},q_{30})$ is the constrained HFB deformation energy,
$\Delta V_{ij}(q_{20},q_{30})$ are the so-called zero-point-energy
corrections, and $B_{ij}(q_{20},q_{30})$ is the inverse of the inertia
tensor $\mathcal{M}_{ij}(q_{20},q_{30})$ associated with the quadrupole
and octupole modes.
In this work, we have taken for $M_{ij}$, instead of the GCM+GOA
inertia tensor, the one deduced from the ATDHF theory with the
Inglis-Belyaev approximation. The reason for this replacement is that
the ATDHF theory appears to give a better account of the nuclear
collective inertia than the GCM one. This question has been extensively
discussed in the literature (see e.g.~\cite{Li99} and references
therein).  \\
The element (ij) of the Inglis-Belyaev inertia tensor can be expressed as:
\be
\displaystyle
\mathcal{M}_{ij}=\sum_{k,l=2,3}(M^{(-1)})^{-1}_{ik}(M^{(-3)})_{kl}(M^{(-1)})^{-1}_{lj}.
\label{mom}
\ee
In Eq.~(\ref{mom}) the moments of order -k are calculated as:
\be
\displaystyle
M^{(-k)}_{ij} = \sum_{\mu \nu}\frac{\bra{\phi(q_{20},q_{30})}|\hat{Q}_{i0}\ket{\mu \nu}
\bra{\mu \nu}|\hat{Q}_{j0}\ket{\phi(q_{20},q_{30})}}{(E_\mu +E_\nu)^k} \ ,
\label{mom2}
\ee
where $\ket{\mu\nu}$ are two quasi-particle states with energies $E_\mu +E_\nu$ built on $\ket{\phi(q_{20},q_{30})}$,
and $\hat{Q}_{i0}$ the
quadrupole/octupole deformation operator
defined in Eqs.~(\ref{q23}) and~(\ref{qq23}), respectively.\\
Let us mention that the collective Hamiltonian $\hat{H}_{coll}$ in Eq.~(\ref{eq6}) is hermitian because: i) all inertia
are real,
and ii) $B_{23}$~=~$B_{32}$. \\
In addition, from Eq.~(\ref{eqtime}), one finds that the collective Hamiltonian and the overlap kernel are even under the
change of $q_{30}$ into $-q_{30}$. Hence, Eq.~(\ref{eq5}) propagates the collective wave function $g(q_{20},q_{30},t)$
without mixing parity components. In particular, if the initial wave function $g(q_{20}, q_{30}, t=0)$ has a good parity
$\pi$, the full time-dependent state Eq.~(\ref{timestate}) will be an eigenstate of $\hat{P}$ with eigenvalue $\pi$. \\
It is important to emphasize at this stage that the approach presented here requires only the use of an effective force.
We recall that the interaction D1S permits one to employ the full HFB theory and consequently to treat the mean field and
the pairing correlations on the same footing at each deformation. Also, the collective Hamiltonian $\hat{H}_{coll}$, as derived
from the GCM procedure, is fully microscopic and relies exclusively on the interaction D1S. Finally let us also add that
the original D1S force is used, which means that no readjustment of the parameters has been made for the application
reported in this paper.  \\
The collective Hamiltonian extracted with our procedure looks like those employed in phenomenological approaches.
However, the form used in the present work directly follows from the TDGCM theory and the GOA ansatz.
We emphasize that the 2 $\times$ 2 inertia tensor depends on the coordinates and is
non-diagonal. Since this situation has not been much studied, numerical methods used to solve Eq.~(\ref{eq5}) are
presented in Section~\ref{sec3}. They differ from the ones previously discussed in ref.~\cite{Be84} because of, first the large
domain of deformation considered here, and second the symmetries of the constraints, which lead to numerical
uncertainties when implementing the previously-used procedures.
%%%%%%%%%%%%%%%%%%%%%%%%%%%%%%%%%%%%%%%%%%%%%%%%%%%%%%%%%%%%%%%%%%%%%%%%%%%%%%%%%%
\section{\label{sec3}Numerical methods}
\subsection{Discretization of the collective variables}
In order to preserve the hermiticity of the collective hamiltonian, the discretization of the collective variables has
been performed by expressing the double integral of the
functional:
\be
\begin{array}{lll}
\displaystyle
F(t)  &  = &  \displaystyle \int \int \; dq_{20} \; dq_{30}\; g^*(q_{20},q_{30},t) \\
      &    &  \displaystyle  (\hat{H}_{coll} \; -i \hbar \; \frac{\partial} {\partial t}) \;g(q_{20},q_{30},t) \ ,
\end{array}
\label{functional}
\ee
with finite differences:
\be
\displaystyle
F(t)=\sum_{ik,jl} g^*(i,k,t) \; K_{ik,jl} \; g(j,l,t) ,
\label{ft}
\ee
and by deriving the discretized equation from the variational principle:
\be
\displaystyle
\frac{\partial F(t)}{\partial g^*(i,k,t)} = 0 .
\ee
In Eq.~(\ref{ft}), K is the symmetric matrix representing $\hat{H}_{coll}$ (whose full expression is given in Appendix A),
and
the labels $i,k$ and $j,l$ correspond to the $q_{20}$ and $q_{30}$ variables through
$q_{20}(i)=(i-1)*\Delta q_{20}$, and
$q_{30}(k)=(k-1)*\Delta q_{30}$.  \\

The time-dependent GCM+GOA equation becomes:
\be
\displaystyle
\sum_{jl} K_{ik,jl} \;g(j,l,t) =\; i \hbar \; \frac{\partial} {\partial t} \; g(i,k,t) .
\label{discret}
\ee

In practice, the two-dimensional discretized form Eq.~(\ref{discret}) has been reduced to a
one-dimensional problem by defining a linear index $m=l+(k-1).l_{max}$, with $l_{max}$ the largest value of $l$
on the grid, which yields:
\be
\displaystyle
\sum_m K_{nm}g_{m}(t) =\; i \hbar \; \frac{\partial}{\partial t} g_{n}(t).
\ee
The resulting 2 $\times$ 2
discretized Hamiltonian matrix $H_{mn}$ is symmetric and the
hermitian character of the kinetic energy operator is preserved. From a numerical point of view, $H_{mn}$ is a sparse matrix.
The corresponding non-zero elements are stored using the "row-indexed storage" method~\cite{Nu86}, an
efficient technique for reducing computing times.

%%%%%%%%%%%%%%%%%%%%%%%%%%%%%%%%%%%%%%%%%%%%%%%%%%%%%%%%%%%%%%%%%%%%%%%%%%%%%%%%%%
\subsection{Time evolution}

In matrix form, the evolution of g between t and t+$\Delta$t can be written:
\be
\displaystyle
g(t+\Delta t) = \displaystyle e^{-i\frac{K \Delta t}{\hbar}} \; g(t).
\label{deltat}
\ee
Using the Crank-Nicholson method~\cite{Nu86,Ca03}, a unitary and stable algorithm, Eq.~(\ref{deltat}) becomes:
\be
\displaystyle
G(t+\Delta t)=\frac{1-i\frac{K \Delta t}{2\hbar}}{1+i\frac{K
\Delta t}{2\hbar}} \; G(t) +O((K \Delta t)^3) \ .
\ee
This equation can be transformed into the linear system:
\be
\displaystyle
(1+i\frac{K\Delta t}{2\hbar}) \; g(t+\Delta t) = \displaystyle (1-i\frac{K\Delta t}{2\hbar})
g(t) \ .
\label{cn}
\ee
In this study, Eq.~(\ref{cn}) is solved by successive iterations until convergence.
The wave function $g(t+\Delta t)$ at time $t+\Delta t$ is determined from the previously known
wave function $g(t)$ at time t as follows:
\be \left \{
\begin{array}{lll}
\displaystyle g^{(n=0)}(t+\Delta t) & = & \displaystyle g(t)  \\
\displaystyle g^{(n+1)}(t+\Delta t) & = & \displaystyle (1-i\frac{K\Delta t}{2\hbar})\;
g(t)     \\
& & \displaystyle -i\frac{K\Delta t}{2\hbar}g^{(n)}(t+\Delta t)   \ .
\end{array}   \right .
\label{timesolve}
\ee
Eqs.~(\ref{timesolve}) are solved in a $q_{20}$ - $q_{30}$ box of finite extension assuming
$g(q_{20},q_{30},t)$~=~0
along the edges of the box. This boundary condition leads to unphysical reflections of the time-dependent wave function
on the $q_{20}$ = $q_{20max}$ edge of the box. In order to eliminate these unphysical reflections
the same technique as that detailed in ref.~\cite{Be91} has been implemented: the wave function is progressively absorbed
in the interior of a rectangular region $q'_{20max} \le q_{20} \le q"_{20max} $beyond the $q_{20} = q_{20max}$ edge
(in the present work $q_{20max}$ = 550 b, $q'_{20max}$ = 800 b and $q"_{20max}$ = 1300 b). Inside this region, the wave
function $g(t)$ is multiplied at each time-step $\Delta t$ by the
function of Woods Saxon structure:
\be
\displaystyle
F(q_{20}) =    \frac{1}{1+exp(-0.015(q_{20}-1150))} \ .
\label{absorb}
\ee
As mentioned in ref.~\cite{Be91}, this technique is similar to adding an imaginary potential $-i\hbar F(q_{20})/\Delta t$
beyond the boundary $q_{20}$ = $q'_{20max}$. Since $\Delta t$ occurs in this imaginary potential, $F(q_{20})$ is
optimized for each time-step.
In the present study the numerical values in Eq.~(\ref{absorb}) have been optimized to avoid reflections
for a time step $\Delta t$~=~1.3~*~$10^{-24}$~s.

The initial wave function $g(t=0)$ is described in terms of
quasi-stationnary vibrational states localized in the first well of the potential energy surface.
The states in question are in fact taken as the eigenstates of a modified two-dimensional $q_{20}$ - $q_{30}$ potential,
where the first fission barrier is extrapolated to large positive values as mentioned in Refs.~\cite{Se93}.
Only the states lying between the top of the inner barrier and 2 MeV above have been considered in the present work.

Fragment mass distributions $Y(A_H)$ are derived by a time-integration of the
flux $\vec{J}(q_{20},q_{30},t) .\vec{n}ds$
of the wave function through scission
at a given fragmentation:
\be
Y(A_H) = \int_{0}^{T} \; dt \; \vec{J}(q_{20},q_{30},t) .\vec{n}ds  \ ,
\label{ya}
\ee
where T is the time for which the time-dependent flux is stabilized along the scission line.
In Eq.~(\ref{ya}), $\vec{n}$ is a vector normal
to the scission line, and $\vec{J}$ is the current defined from the continuity equation:
\be
\frac{d}{dt}|g(q_{20},q_{30},t)|^2=-div \vec{J}(q_{20},q_{30},t).
\ee
The current $\vec{J}=(J_2,J_3)$ as calculated with the collective Hamiltonian defined in eq.~(\ref{eq6}) takes the form:
\be
\begin{array}{ccc}
J_2(q_{20},q_{30},t) & = & \displaystyle \frac{\hbar}{2 i}(g^* B_{22} \frac{\partial g}{\partial q_{20}} -g
B_{22} \frac{\partial g^*}{\partial q_{20}} \\
& &  \displaystyle +g^* B_{23} \frac{\partial g}{\partial q_{30}} -gB_{23} \frac{\partial g^*}{\partial q_{30}}) \ ,\\
& & \\
J_3(q_{20},q_{30},t) & = &  \displaystyle \frac{\hbar}{2 i}(g^* B_{33} \frac{\partial g}{\partial q_{30}} -g
B_{33} \frac{\partial g^*}{\partial q_{30}}   \\
& &  \displaystyle +g^* B_{32} \frac{\partial g}{\partial q_{20}} -gB_{32} \frac{\partial g^*}{\partial q_{20}}) \ .
\end{array}
\label{eq26}
\ee
Expression~(\ref{eq26}) reveals in particular that the component of the current in one direction involves the gradients in all
directions. This observation will be used in section~\ref{sec5}, where we discuss the contributions of interference
terms between components of different parities in the initial state.

%%%%%%%%%%%%%%%%%%%%%%%%%%%%%%%%%%%%%%%%%%%%%%%%%%%%%%%%%%%%%%%%%%%%%%%%%%%%%%%%%%
\section{\label{sec4}Static results}
\subsection{\label{sec3sub1}Potential energy surface}
\begin{figure}[htb]
\begin{center}
%\vskip -4 cm
%\hskip -7 cm
\includegraphics[scale=0.3,angle=-90]{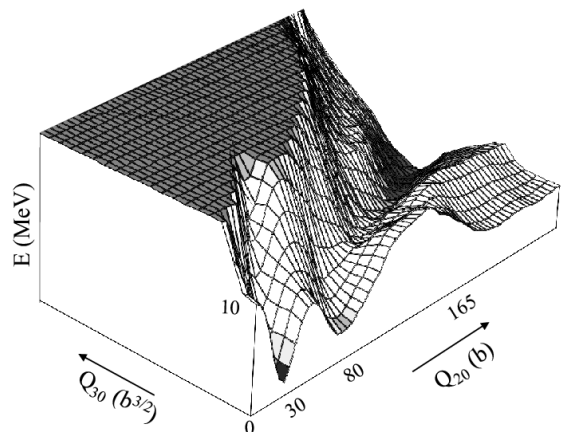}
%\vskip -0.5 cm
\caption{HFB potential energy surface as a function of $\rm q_{20}$ and $\rm q_{30}$ collective variables in $^{238}$U.}
\label{figg1}
\end{center}
\end{figure}

HFB calculations have been performed for $^{238}$U with constraints on both the quadrupole and octupole
moments, using the mesh sizes $\Delta q_{20}$~=~5~-~10~b and $\Delta q_{30}$~=~2~-~4 b$^{3/2}$. The range of
investigation extends from spherical shapes ($q_{20}$~=~0 b) up to elongations of the exit points, which vary from
$q_{20max}^a$~=~320~b for the most asymmetric fission ($q_{30}$ = 44 b$^{3/2}$) up to
$q_{20max}^s$~=~550~b for symmetric fragmentation ($q_{30}$ =~0~b$^{3/2}$). For each value of the quadrupole moment,
the HFB calculations have been restricted to solutions whose excitation energies are at most 30 MeV above the ground-state.
For values of $q_{20}$ near scission, this condition leads to a maximum value of $q_{30}$~=~120~b$^{3/2}$.
HFB solutions for 120~$<$~$q_{30}$~$<$~200 b$^{3/2}$ have been extrapolated.  \\
Fig.~\ref{figg1} shows the most significant part of the HFB potential energy surface as a function of the quadrupole
and octupole moments. For practical reasons, the domain of the plot is restricted to 0 $<$ $q_{20}$ $<$ 320 b and
0 $<$ $q_{30}$ $<$ 72 b$^{3/2}$ and energies are truncated to 25 MeV. As expected in this actinide nucleus, the
ground-state is found to be deformed with $q_{20}$ $\approx$ 30 b and a super-deformed minimum appears for an elongation
close to $q_{20}$ = 80 b. Beyond this second well, two valleys appear. They are separated by a ridge for well-elongated
shapes and lead either to the symmetric or to the most probable asymmetric fragmentations.

For each asymmetry, the determination of scission configurations is made by increasing the elongation step by step:
the constrained HFB wave function at a given $q_{20}$ is generated from a previous solution at a slightly lower
elongation while keeping $q_{30}$ fixed.
This method relies on the scission mechanism studied in~\cite{Be84}. It is assumed that scission occurs for a given value of
$q_{30}$ when the system falls from the so-called "fission valley" to the "fusion valley" describing well-separated
fragments.
%\newpage

\begin{figure}[htb]
\begin{center}
\includegraphics[width=8.5cm,angle=0]{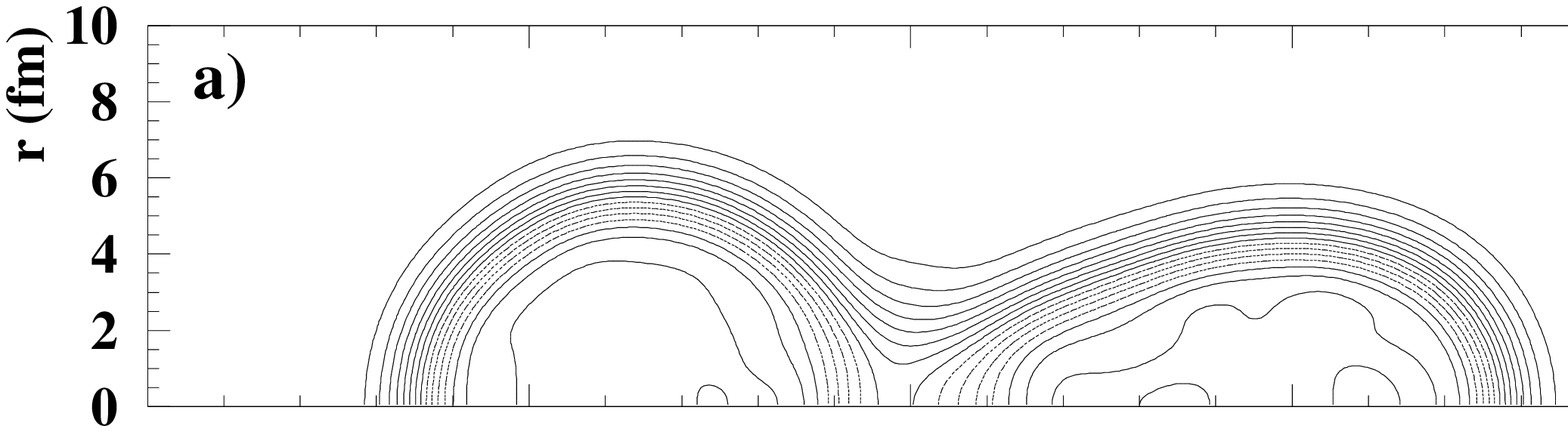}
\vglue -6.7 truecm
\includegraphics[width=8.5cm,angle=0]{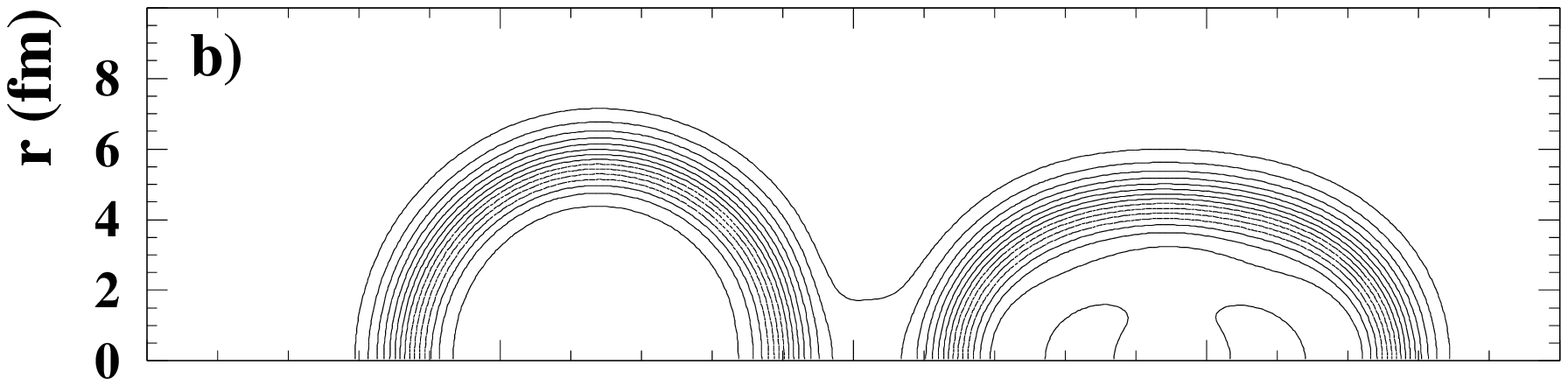}
\vglue -1.8 truecm
\includegraphics[width=8.5cm,angle=0]{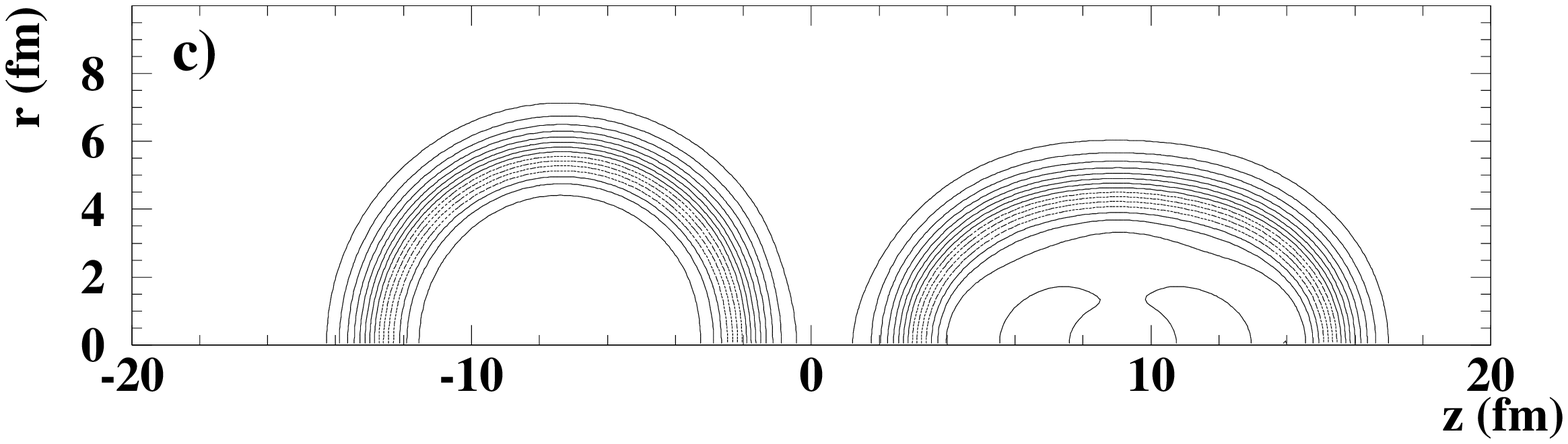}
\end{center}
\vglue -0.5 truecm
\caption{Proton plus neutron density contours at a given asymmetry $q_{30}$ = 44 b$^{3/2}$ for different elongations
a) $q_{20}$ = 310 b, b) $q_{20}$ = 320 b and c) $q_{20}$ = 330 b.
Contour lines are separated by 0.01 nucleons fm$^{-3}$. }
\label{densite}
\end{figure}

The main criterion used to define exit points and to separate pre-
and post- scission configurations is obtained by looking at the nucleon density in the neck: we consider that the system
is composed of two fragments when the density in the neck is less than 0.01 nucleon/fm$^3$. This is illustrated in
Fig.~\ref{densite}, where density contours are plotted for a given asymmetry $q_{30}$ = 44 $b^{3/2}$ and an increasing
elongation. Contour lines are separated by 0.01 nucleon/fm$^3$. Figures~\ref{densite}(a) and~\ref{densite}(b) correspond
to pre-scission configurations and figure~\ref{densite}(c) to a post-scission one. Let us note that the two criteria
described
in Ref.~\cite{Be84} are also satisfied: a~$\simeq$ 15 MeV drop in the energy of the total system, and
a~$\simeq$~30$\%$ decrease of the
hexadecapole moment are observed when scission occurs.

%\newpage
It is worth pointing out that the constrained HFB method does not impose an {\it a priori} shape to the fissioning
system. All types of deformations which are not imposed take the values that minimize the total nuclear energy with both
the nuclear mean field and pairing field determined self-consistently. Results concerning fragment deformations at scission
will be presented in a forthcoming publication.

\begin{figure}[htb]
\begin{center}
\includegraphics[scale=0.30]{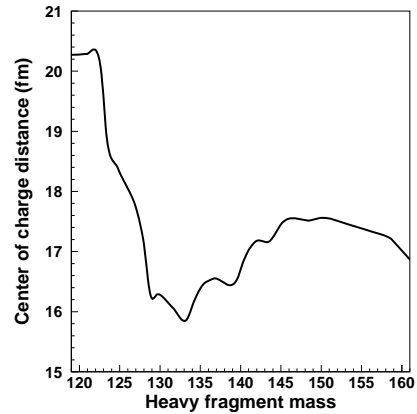}
\caption{Distance between the centers of charge of the fragments as a function of the heavy fragment mass.}
\label{figd}

\end{center}
\end{figure}

%\newpage
Near the exit points, the z-location of the neck, $z_{neck}$, is determined as the z-value for which the nucleon density
integrated over $r$ is minimum. Properties of the fragments, such as their masses and their charges, their deformations
and the distance between their centers of charge are calculated from integrations in the left and right half-spaces on
either sides of the z = $z_{neck}$ plane.
As an example, the distance d between the centers of charge of the fragments is plotted in Fig.~\ref{figd} as a function
of the heavy fragment mass. It is found to be maximum for $A_H$~=~119 with d~=~20.27 fm and minimum for $A_H$~=~134 with
d~=~15.88 fm. Precise values of this fragment center of charge distance are crucial because they govern the Total Kinetic
Energy (TKE)
distribution, as discussed in Section~\ref{sec3sub5}.
As a test, we have checked that the analytical relations in Eq.~(\ref{q23b}) are fulfilled.
%%%%%%%%%%%%%%%%%%%%%%%%%%%%%%%%%%%%%%%%%%%%%%%%%%%%%%%%%%%%%%%%%%%%%%%%%%%%%%%%%%
\subsection{\label{sec3sub3}Pairing correlations}
\begin{figure}[htb]
\begin{center}
\includegraphics[scale=0.31,angle=-90]{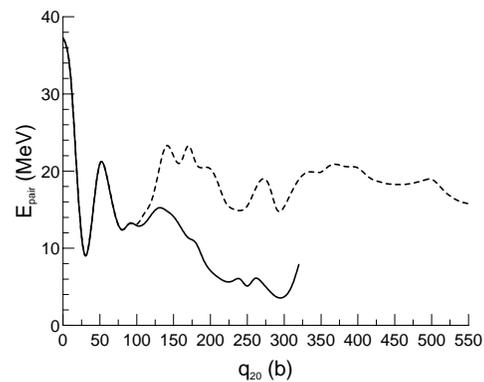}
\caption{Pairing energy as a function of $q_{20}$ along the asymmetric (solid line) and the symmetric
(dashed line) fission paths in $^{238}$U.}
\label{fig2}
\end{center}
\end{figure}

Fig.~\ref{fig2} shows the pairing energy
$E_{pair}=\frac{1}{2}Tr(\Delta \kappa)$, where $\Delta$ and
$\kappa$ are the pairing field and the pairing tensor, respectively. We clearly see that pairing is not constant as a
function of elongation. As expected, minima are found inside the wells and maxima at the top of barriers. Furthermore,
the total pairing energy, $E_{pair}$, is predicted to be lowest in the asymmetric valley ($E_{pair} \approx $ 6 MeV) and
much larger in the symmetric one ($E_{pair}$ $>$ 15MeV). These variations of the pairing correlations are very
important
since they strongly influence both the collective flux and the occurrence of intrinsic excitations, as is now explained.

\begin{figure}[htb]
\begin{center}
\includegraphics[scale=0.31,angle=-90]{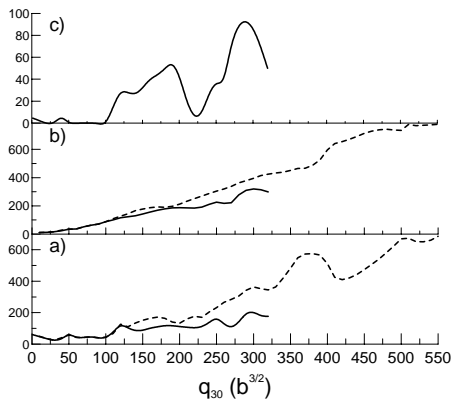}
\caption{Components of the inertia tensor a) $B_{22}$ (in MeV~$b^2$~$\hbar^2$) b) $B_{33}$ (in MeV~$b^3$~$\hbar^2$)
c) $B_{23}$ (in MeV~$b^{5/2}$~$\hbar^2$) as functions of $q_{20}$ along the
asymmetric (solid line) and the symmetric
(dashed line) fission paths}
\label{fig3}
\end{center}
\end{figure}

First, the collective inertia is known to be very sensitive to pairing correlations. The three components $B_{22}$,
$B_{33}$ and $B_{23}$ of the inertia tensor in Eq.~(\ref{eq6}) are plotted in
Figs.~\ref{fig3}(a),~\ref{fig3}(b) and~\ref{fig3}(c), respectively, as functions of the elongation along the symmetric
(dotted line) and asymmetric (solid line) paths. The two components $B_{22}$ and $B_{33}$ are found to be larger in the
symmetric valley than in the asymmetric one (up to a factor of two at large elongation). Furthermore,
whereas the non-diagonal inertia component $B_{23}$ is zero for $q_{30}~$=~0 by definition, $B_{23}$ is found to be
non negligible as soon as the system spreads widely in the asymmetric valley. The coupling brought by $B_{23}$
between the $q_{20}~$ and $q_{30}~$ modes indicates that, as time evolves,
the two collective degrees of freedom exchange energy, which will affect, among other things, the
kinetics of the fission process.

\begin{figure}[htb]
\begin{center}
\includegraphics[scale=0.31,angle=-90]{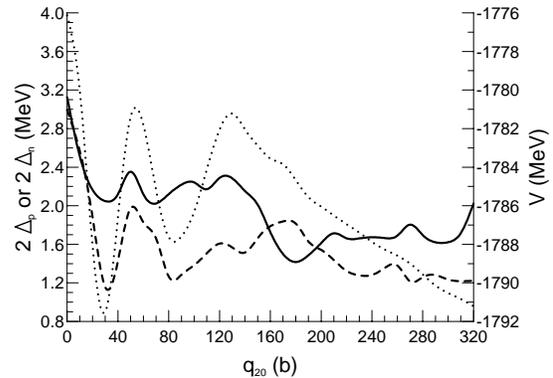}
\caption{Twice the proton (solid curve) and neutron (dashed curve) lowest quasi particle energies and potential energy
(dotted curve)
along the asymmetric path as functions of the elongation in barns.}
\label{fig4}
\end{center}
\end{figure}

Second, pairing correlations characterize the amount of superfluidity of the collective flux and the onset of
dissipation, in particular
between the saddle point and the exit point. In the HFB approach, dissipation requires the creation of two
quasiparticle excitations, that is a transfer of energy from the collective motion at least equal to 2$\Delta$,
where $\Delta$ is the energy necessary to break a correlated pair. One expects that small values of $\Delta$ will
favor "dissipation".
However, the excitation of the intrinsic structure also depends on
the coupling between collective
and intrinsic degrees of freedom, which is largely unknown.
For this reason, the question of dissipation effects
will be addressed in future work.
%\newpage
The proton and neutron gaps $2\Delta_p$, $2\Delta_n$ are plotted in Fig.~\ref{fig4} as functions of elongation along
the asymmetric path. The corresponding potential energy curve is also plotted (dotted curve) to guide the eyes. For
proton pairing correlations, we find $2\Delta_p$ = 2.3 MeV at the top of the second barrier. This value appears to be in good
agreement with experimental data~\cite{Po93,Po94,Vi00}. As a matter of fact, manifestations of proton pair breaking
are observed in $^{238}$U
and $^{239}$U nuclei for an excitation energy of 2.3 MeV above the barrier: first the proton odd-even effect observed in
the fragment mass distributions
decreases
exponentially for excitation energy slightly higher than 2.3 MeV~\cite{Po93} and second, the total kinetic energy drops
suddenly ~\cite{Po94,Vi00}.

In Fig.~\ref{fig4} we also see that the proton gap decreases rapidly during the first part of the descent beyond the saddle
point, for instance 2$\Delta_p$ = 1.4 MeV for $q_{20}$~=~180~b.
However, experimental facts show that increasing the excitation energy from 0 to 2.2 MeV above the barrier does not modify
the proton odd-even effect. From our point of view, this could be an indication that no proton pairs are broken during
the descent from saddle to scission in low energy fission for excitation energies below 2.2 MeV above the barrier. Our
interpretation is that the excitation energy supplied during the descent is shared among the collective degrees of freedom
and not among intrinsic excitations. This experimental observation gives us some confidence that the neglect of the
coupling between collective and intrinsic degrees of freedom is a reasonable approximation to start with in low energy
fission.

Finally, no strong odd-even neutron effects are observed for the fragment mass distributions measured in the photofission of
$^{238}$U, regardless of excitation energy~\cite{Po94}. In our calculations the neutron pairing gap is found to be
much lower than the proton one, except for 160 $<$ $q_{20}$ $<$ 190 b , as displayed in Fig.~\ref{fig4}. At the top
of the second barrier the neutron gap is only 2$\Delta_n$ = 1.6 MeV. This tends to indicate that neutron pairs are more
likely to be broken than proton ones in the even-even $^{238}$U nucleus at low excitation energy. But no definite
comparison with experimental data can be made since a precise knowledge of the neutron number of the fission fragments
is made extremely difficult by the neutron evaporation.
All these remarks concerning pairing correlations {\it a posteriori} illustrate the fact that pairing correlations play
an essential role and that they should be introduced in dynamical studies of fission.
%%%%%%%%%%%%%%%%%%%%%%%%%%%%%%%%%%%%%%%%%%%%%%%%%%%%%%%%%%%%%%%%%%%%%%%%%%%%%%%%%%
\subsection{\label{sec3sub5}Total Kinetic Energy distribution}
\begin{figure}[htb]
\begin{center}
\includegraphics[scale=0.4]{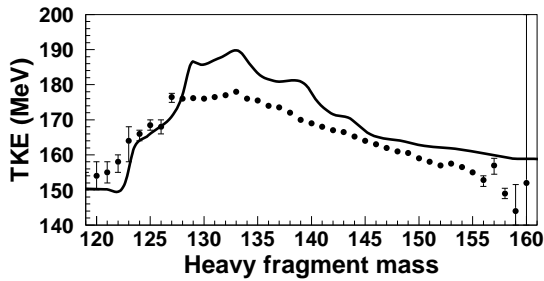}
%\vskip -4 cm
\caption{Total kinetic energy distributions as functions of the heavy fragment mass.
Dots indicate experimental data~\cite{Po94} and the continuous line present predictions.}
\label{fig6}
\end{center}
\end{figure}

As a first estimate, the total kinetic energy $TKE$ of the fragments can be roughly calculated as the Coulomb
potential energy $TKE =\frac{Z_HZ_Le^2}{d}$, with d the distance between the centers of charge of the fragments at
scission. \\
Theoretical values calculated along the scission line
are shown in Fig.~\ref{fig6} as a function of $A_H$, the heavy fragment mass. They are compared to experimental data
obtained from the photofission of $^{238}$U using 6.2 MeV bremsstrahlung
$\gamma$~rays, corresponding to an excitation energy close to the inner fission barrier height~\cite{Po94}.
We first notice that the general trend of the distribution is rather well reproduced, with a dip at $A_H$ = 119 and a peak
for $A_H$ = 134. Symmetric and asymmetric wings are surprisingly close to experimental data.
The agreement indicates that our microscopic approach, together with the prescription explained in Section~\ref{sec3sub1}
is able to give a realistic description of scission configurations.
The main difference with experimental data occurs in the region of the most probable asymmetric fission where the theoretical
results overestimate TKE values by $\simeq$ 6$\%$.
This discrepancy mainly comes from the fact that the nuclear contribution entering the mutual energy between the two
fragments is not strictly zero for the corresponding scission configurations. Furthermore, the -- attractive --
exchange Coulomb energy between the fragments has been neglected. These two effects could lead to a decrease of TKE
values that may reach 10~-~15~MeV.

%%%%%%%%%%%%%%%%%%%%%%%%%%%%%%%%%%%%%%%%%%%%%%%%%%%%%%%%%%%%%%%%%%%%%%%%%%%%%%%%%%
\subsection{\label{static}"One-dimensional" fragment mass distribution}

As a first approximation, mass distributions can be derived using the
fragmentation model detailed in ref.~\cite{Li73}.
Namely, collective stationary vibrations along the sole mass-asymmetry
degree of freedom for nuclear configurations just before scission are
studied.
The probability of occurrence of a mass asymmetry ($A_H$,~$A_L$)
corresponding to a value $q_{30}$ of the octupole moment is then taken
as:
\be
Y(A_H,A_L) = |\Psi_0^{+1}(q_{30})|^2   \ ,
\label{yield}
\ee
where $\Psi_0^{+1}$ is the positive parity eigenstate with lowest
energy of the one-dimension collective Hamiltonian
$\hat{H'}_{coll}$ in the $q_{30}$ variable:
\begin{equation}
\begin{array}{lll}
\displaystyle \hat{H'}_{coll} (q_{30},\frac{\partial}{\partial q_{30}})&  = &\displaystyle
   -\frac{\hbar^2}{2}\frac{\partial}
{\partial q_{30}}\frac{1}{\mathcal{M}_{3}(q_{30})}
\frac{\partial}{\partial q_{30}} \\
&    &       \\
&    & \displaystyle   +V(q_{30})-  \Delta V_{3}(q_{30}) \ .
\end{array}
\label{q31d}
\end{equation}
Here, $V(q_{30})$ is the HFB deformation energy along the scission line $q_{20}$~=~$q_{20}^s$~=~$f(q_{30})$,
$\mathcal{M}_3(q_{30})$ is the collective inertia,
and $\Delta V_{3}(q_{30})$ the zero-point-energy correction (ZPE).
The Hamiltonian of Eq.~(\ref{q31d}) is derived from the usual GOA
reduction of the one-dimensional Hill-Wheeler stationary equation
obtained by taking for the generator coordinate the curvilinear
abscissa $s(q_{20},q_{30})$ along the scission line.
A change of variable is then performed in order to express all quantities
as functions of $q_{30}$.
It is easy to show that the inertia and ZPE correction appearing in
Eq.~(\ref{q31d}) are related to those entering the full
two-dimensional Hamiltonian (\ref{eq6}):
\be
\displaystyle
\mathcal{M}_3(q_{30})  =  (\frac{df}{dq_{30}})^2 \, M_{22} +2 (\frac{df}{dq_{30}}) \, M_{23} +M_{33},
\ee
\be
\displaystyle
\Delta V_{3}(q_{30})   =  \frac{\mathcal{G}_3(q_{30})}{2\mathcal{M}_3(q_{30})},
\label{iner}
\ee
with
\be
\mathcal{G}_3(q_{30}) =(\frac{df}{dq_{30}})^2 \, G_{22} +2 (\frac{df}{dq_{30}}) \, G_{23} +G_{33} ,
\ee
where $M_{ij}$ are the inertia defined in Eq.~(\ref{mom}) and $G_{ij}$ the components of the overlap tensor calculated in
the cranking approximation using the moments of Eq.~(\ref{mom2}).

Clearly, the model based on the Hamiltonian (\ref{q31d}) amounts to
ignore all the effects of the dynamics along the elongation degree of
freedom from the first well to scission.
We call the mass distribution obtained in this way a "one-dimensional" mass distribution.

\begin{figure}[htb]
\begin{center}
\includegraphics[scale=0.31,angle=-90]{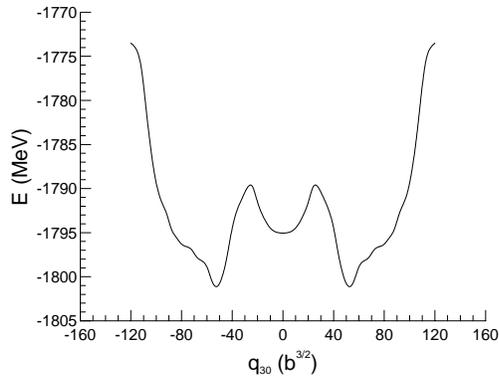}
\caption{Potential energy along the scission line as a function of the octupole moment.}
\label{statpot}
\end{center}
\end{figure}

The HFB potential energy $V(q_{30})$ calculated along the scission line is plotted in
Fig.~\ref{statpot} as a function of the octupole moment. The lowest energy is obtained for $q_{30}$ = $\pm$ 44 b,
corresponding to the most probable fission. A secondary minimum is found for $q_{30}$~=~0~b. These two wells are
separated by a 11 MeV high barrier.

\begin{figure}[htb]
\begin{center}
\includegraphics[scale=0.31,angle=-90]{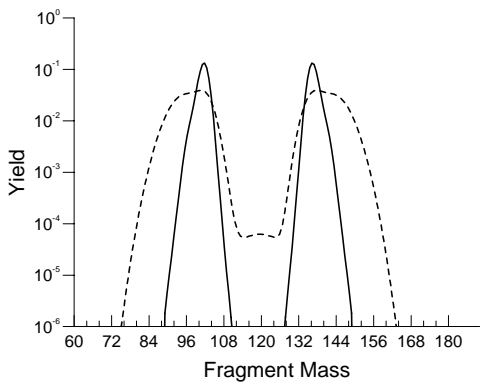}
\caption{Comparison between the "one-dimensional" mass fragment distribution obtained from~(\ref{yield}) (solid line), and
the Wahl evaluation (dashed line)~\cite{Wa02}.}
\label{statique}
\end{center}
\end{figure}

%\newpage
"One-dimensional" distributions are shown in Fig.~\ref{statique} where the mass yield Eq.~(\ref{yield}) is
plotted (solid line)
together with the Wahl evaluation (dashed line) for 46 keV neutron induced fission on $^{237}$U~\cite{Wa02}. The maxima
of the theoretical curve occur at $A_H$ = 134, $A_L$ = 94 values corresponding to the minima of the potential energy
along the scission line. The fact that the experimental curve maxima lie close to these values indicate that the most
probable fragmentation is due essentially to the properties of the potential energy surface at scission, i.e mainly to
shell effects in the nascent fragments.
However, the "one-dimensional" approach does not reproduce neither the experimental peak-to-valley ratio nor
the experimental widths of the distributions - the theoretical
widths are twice smaller than the Wahl evaluated ones-.

One must note however that only the solution of Eq.~(\ref{q31d}) with lowest energy $\Psi_0^{+1}$ has so far been
considered. This is certainly an oversimplifying assumption since, the wave function describing the collective
evolution of the nucleus will undoubtedly possess a more complicated structure at the time of reaching the
scission line. In particular,
as mentioned in ref.~\cite{Ma74}, $n>0$ states $\Psi_n^{\pi}$, solutions of Eq.~(\ref{q31d}), may become
excited due to the interaction
between $q_{20}$ and $q_{30}$ degrees of freedom.
Let us mention that, in Ref~\cite{Ma74} the population of the eigenstates has been assumed to follow a Boltzmann law governed
by a temperature
parameter. In Ref.~\cite{Ma76}, the elongation degree of freedom has been introduced using a classical approximation.

\begin{figure}[htb]
\begin{center}
\vskip -1cm
\includegraphics[scale=0.31, angle=-90]{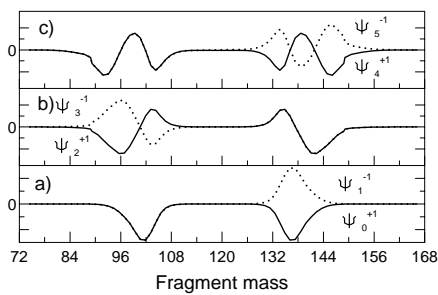}
\caption{Amplitudes of the first six collective states in
the asymmetry variable at scission as functions of the fragment mass. Positive (negative) parity states are plotted
in solid (dotted) lines. Excitation energies, measured from the lowest energy state, are a)~E~=~0~MeV, b)~E~=~3.09~MeV
and, c)~E~=~4.76~MeV.}
\label{staticfig}
\end{center}
\end{figure}

Amplitudes of the first six collective states $\Psi_n^{\pi}$, n~=~0~...~5, solutions of Eq.~(\ref{q31d}) are displayed
in Fig.~\ref{staticfig} as functions of the fragment mass.
As is well-known, such $\Psi_n^{\pi}$ states are eigenstates of the parity operator $\Pi$ with eigenvalues ${\pi}$.
Positive and negative parity states are plotted in solid and dotted lines, respectively. Each pair of $\pi$ = +1 and
$\pi$ = -1 levels is degenerate in energy because the potential is symmetric with respect to the
$q_{30}$~$\rightarrow$~-$q_{30}$ transformation, and because the barrier between the two asymmetric wells is high (11 MeV)
(see Fig.~\ref{statpot}). We observe in Fig.~\ref{staticfig} that
the excited states are more spread over mass than the ground state. For example, the wave functions $\Psi_4^{+1}$ and
$\Psi_5^{-1}$
displayed in Fig.~\ref{staticfig}c) display non-zero values up to $A_H \approx$~156 whereas the lowest energy wave functions
$\Psi_0^{+1}$ and $\Psi_1^{-1}$ in Fig.~\ref{staticfig}a) are localized in the domain 132~$<$~$A_H$~$<$ 144. Therefore,
we can expect that the introduction of these excited states in the definition of the mass yield Eq.~(\ref{yield}) will
broaden the mass distribution in the asymmetric region.

%%%%%%%%%%%%%%%%%%%%%%%%%%%%%%%%%%%%%%%%%%%%%%%%%%%%%%%%%%%%%%%%%%%%%%%%%%%%%%%%%%
\section{\label{sec5}Dynamical results}
%%%%%%%%%%%%%%%%%%%%%%%%%%%%%%%%%%%%%%%%%%%%%%%%%%%%%%%%%%%%%%%%%%%%%%%%%%%%%%%%%%
\subsection{Initial states}
The time-dependent evolution of the system has been calculated using different initial conditions in the first
well of the potential energy surface.
Calculations have been performed from t = 0 up to maximum times for which the flux of
the time-dependent collective wave function along the scission line has become stabilized.

We first discuss the effect of the structure of the initial state on the mass distribution.
In order to define the initial conditions we imagine that the nucleus is a compound system described in terms of
complicated quasi-stationnary states which decay into various channels (neutron and $\gamma$-ray emission and fission).
In the case
of the even-even K~=~0 $^{238}$U nucleus studied here, we assume that states which decay through fission can be
described by the simple form:
\be
\displaystyle
\ket{\Psi_{P,K=0,I,M}}= (2\pi)^{-\frac{1}{2}} Y_{I,M} (\Omega) \; \int dq \; f_n^{\pi} (q,t=0) \; \ket{\phi(q)} ,
\label{wftot}
\ee
where $Y^M_I(\Omega)$ are spherical harmonics and $\Omega$ the Euler angles relating the intrinsic axes of the
nucleus to the laboratory frame of reference.

\begin{figure}[htb]
\begin{center}
\vskip -0.8 cm
\includegraphics[scale=0.3,angle=-90]{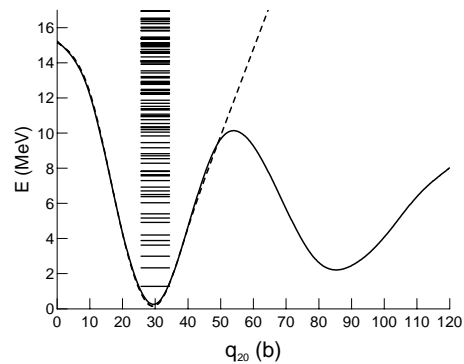}
\caption{Potential curves $V(q_{20},q_{30}=0)$ and $V'(q_{20},q_{30}=0)$ including zero-point-energy corrections
(continuous
and dotted lines, respectively), and
collective eigenstates of the modified $V'(q_{20},q_{30}=0)$ potential (horizontal segments).}
\label{potVVprim}
\end{center}
\end{figure}

The parity quantum number P is related to the intrinsic parity $\pi$ by the following relation~:
\be
P=\pi(-1)^I  ,
\label{parite}
\ee
where I is the spin of the fissioning system. In Eq.~(\ref{wftot}), q is the set of all relevant nuclear
collective deformations which,
in the present work (see Eq.~(\ref{eq1})) is restricted to $(q_{20},q_{30})$.
 \\
As already mentioned in Section~\ref{sec2}, initial states $g_n^{\pi}(q_{20},q_{30},t=0)$
(related to the $f_n^{\pi}(q_{20},q_{30},t=0)$ functions as in
Eq.~(\ref{timestate})) are taken as eigenstates of the modified
two-dimensional first well $V'(q_{20},q_{30})$, where the potential has
been extrapolated at large deformations as shown in
Fig.~\ref{potVVprim}. They are solutions of the equation:
\be
\displaystyle
\hat{H'}_{coll} \;g_n^{\pi}(q_{20},q_{30},t=0) = E^{\pi}_n \;g_n^{\pi}(q_{20},q_{30},t=0) \ ,
\ee
where $\hat{H'}_{coll}$ is the Hamiltonian defined in Eq.~(\ref{eq6}) with $V(q_{20},q_{30})$ replaced
by $V'(q_{20},q_{30})$.

Because $H'_{coll}(q_{20},q_{30},\frac{\partial}{\partial q_{20}},\frac{\partial}{\partial q_{30}})$ =
$H'_{coll}(q_{20},-q_{30},\frac{\partial}{\partial q_{20}},-\frac{\partial}{\partial q_{30}})$ these initial states
can be chosen as
eigenstates of the parity operator with eigenvalues $\pi$~=~$\pm$~1~\cite{Me95}:
\be
\displaystyle
g_n^{\pi}(q_{20},-q_{30},t=0) = \pi \; g_n^{\pi}(q_{20},q_{30},t=0) \ .
\ee
The potential curves $V(q_{20},q_{30}=0)$ and $V'(q_{20},q_{30}=0)$ including zero-point-energy corrections
are displayed in Fig.~\ref{potVVprim}.
The eigen-energies of $H'_{coll}$ are also shown.
Excitation energies of the compound nucleus in the interval
$[$~$B_{I}$~,~$B_{I}$~+2.5~MeV~$]$ will be considered in this work focussing on
low-energy fission, where $B_{I}$ is the first barrier height. In this energy range,
the mean-level spacing is found to be
around 130 keV. Therefore, 19 states are possible initial candidates for our dynamical calculations.
All these states are located above the outer symmetric saddle point but below the outer asymmetric one.
They correspond to multi- quadrupole and octupole phonons, and have different components along the $q_{30}$ and $q_{20}$
directions. Significant effects on the fragment mass distributions are mainly due to the parity of
the initial states.
%\newpage

\begin{figure}[htb]
\begin{center}
\includegraphics[scale=0.3,angle=-90]{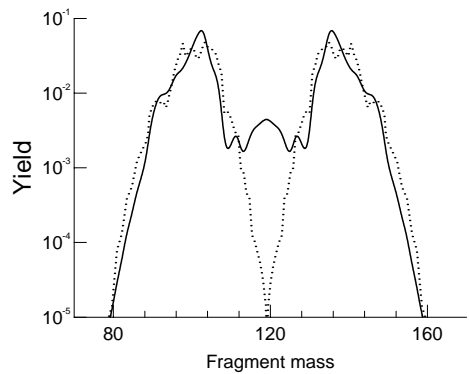}
\caption{Fragment mass distributions obtained for initial states having positive parity (solid line) and
negative parity (dotted line).}

\label{plusmoins}
\end{center}
\end{figure}

In Fig.~\ref{plusmoins}, fragment mass distributions, calculated with
formula~(\ref{ya}) and initial states of definite parity, are plotted
separately. The solid and dotted curves correspond to initial states
whose intrinsic parity is positive or negative exclusively. They are
located at 2.4 MeV and 2.3 MeV above the first barrier, respectively.

We see that the main difference between the two results is the peak-to-valley ratio which is of the order of 50
for the positive parity initial state and infinite for the negative one. The fact that no symmetric fission is found
when the initial state has a negative parity is due to the fact that $g_n^{-1}(q_{20},q_{30}=0,t)$~=~0 if $\pi$~=~-1
for any time t.
As a consequence, the flux of the wave function through the scission line at $q_{30}$~=~0 vanishes.  \\
In the applications presented below we use initial states that don't have a definite parity. As we observed at the end of
section~\ref{sec3} there are interferences between states of different parities in the calculation of the flux.
However, due to the symmetries of the inertia tensor, these interferences do not contribute to the symmetric
mass fragmentation.
>From the previous discussion we infer that our predictions of symmetric fission will be affected by the proportions of
collective states with negative and postive intrinsic parity. In order to get an estimate of these proportions we assume
that they are the same as in the compound sytem n+$^{237}$U. More precisely, by using Eq.~(\ref{parite}) we define
fission cross
sections, $\sigma$($\pi$~=~+1,E) and $\sigma$($\pi$~=~-1,E) corresponding to components of intrinsic parity in the compound
system through the relations:
\be
\begin{array}{lll}
\sigma(\pi = -1,E)& =  \displaystyle \sum_{I=2p,P=-1}&\sigma_{CN}(P,I,E) \;  P_f(P,I,E) \\
                 & \displaystyle + \sum_{I=2p+1,P=+1}& \sigma_{CN}(P,I,E) \;  P_f(P,I,E) \ , \\
              & & \\
\sigma(\pi = +1,E)& =  \displaystyle \sum_{I=2p,P=+1}&\sigma_{CN}(P,I,E) \; P_f(P,I,E) \\
                 & \displaystyle + \sum_{I=2p+1,P=-1} &\sigma_{CN}(P,I,E) \; P_f(P,I,E)  \ ,
\end{array}
\ee
where E, P, $\sigma_{CN}(P,I,E)$ and $P_f(P,I,E)$ are the energy, the parity (defined in Eq.~(\ref{parite})),
the formation cross section and the
fission probability of the compound nucleus, respectively.

Formation cross sections have been calculated using the Hauser-Feschbach theory with the optical potential model of
Ref.~\cite{optic} and fission probabilities
have been deduced from a statistical model calculation~\cite{WKB}.

Then, we define probabilities by the following fractions:
\be
\begin{array}{lll}
p^-(E) & = & \displaystyle \frac{\sigma(\pi = -1,E)}{\sigma(\pi = -1,E)+\sigma(\pi = +1,E)}  \ ,\\
              & & \\
p^+(E) & = & \displaystyle \frac{\sigma(\pi = +1,E)}{\sigma(\pi = -1,E)+\sigma(\pi = +1,E)}  \ .
\end{array}
\ee

They represent the population of states in the compound system which have a given intrinsic parity and which decay to
fission. It is with the help of these probabilities that we determine the mixing of parities in the initial states.
Numerical values for the reaction n+$^{237}$U are given in Table~\ref{percent} for two excitation energies as
measured from the top of the first barrier.
\renewcommand{\arraystretch}{1.5}
\begin{table}[htb]
\begin{tabular}{|l|l|l|}     \hline
$\; \;\;$E (MeV)       $\; \;\;$  & $\; \;\;$ 1.1 $\; \;\;$     & $\; \;\;$ 2.4  $\; \;\;$  \\  \hline
$\; \;\;$$p^+(E)$ $\%$ $\; \;\;$  & $\; \;\;$ 77  $\; \;\;$     & $\; \;\;$ 54  $\; \;\;$    \\  \hline
$\; \;\;$$p^-(E)$ $\%$ $\; \;\;$  & $\; \;\;$ 23  $\; \;\;$     & $\; \;\;$ 46   $\; \;\;$   \\  \hline
\end{tabular}
\vskip 0.7 cm
\caption{Percentages of positive and negative intrinsic parity states populated in the compound nucleus $^{238}$U
by the n+$^{237}$U reaction for two excitation energies.}
\label{percent}
\end{table}

\begin{figure}[htb]
\begin{center}
\includegraphics[scale=0.3,angle=-90]{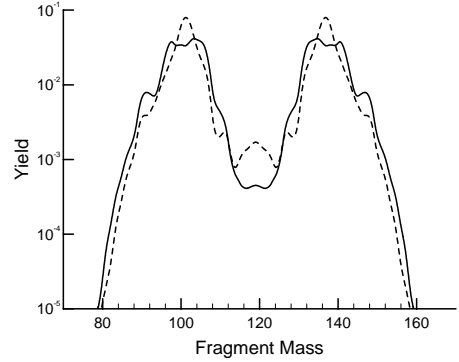}
\caption{Fragment mass distributions obtained for two initial states. Solid line: E~=~2.4~MeV with $p^+$~=54~$\%$
and $p^-$~=46~$\%$. Dashed line: E~=~1.1~MeV with $p^+$~=~77~$\%$ and $p^-$~=~23~$\%$.}
\label{n1n2}
\end{center}
\end{figure}

Mass distributions obtained for these two energies are displayed in Fig.~\ref{n1n2}: the solid and dashed lines
correspond to the two energies, 2.4 MeV and 1.1 MeV respectively.

\begin{figure}[htb]
\begin{center}
\includegraphics[scale=0.3,angle=-90]{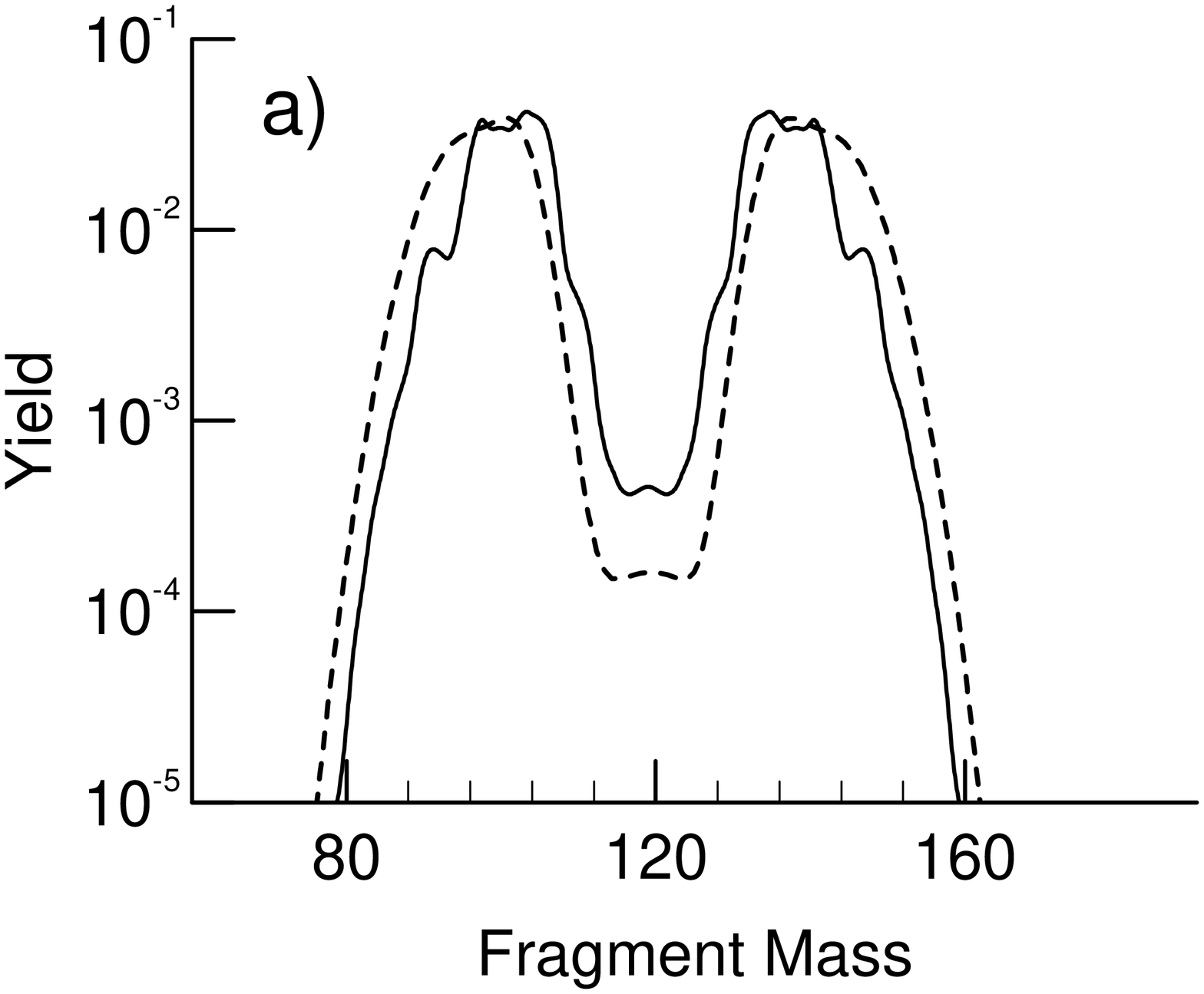}
\includegraphics[scale=0.3,angle=-90]{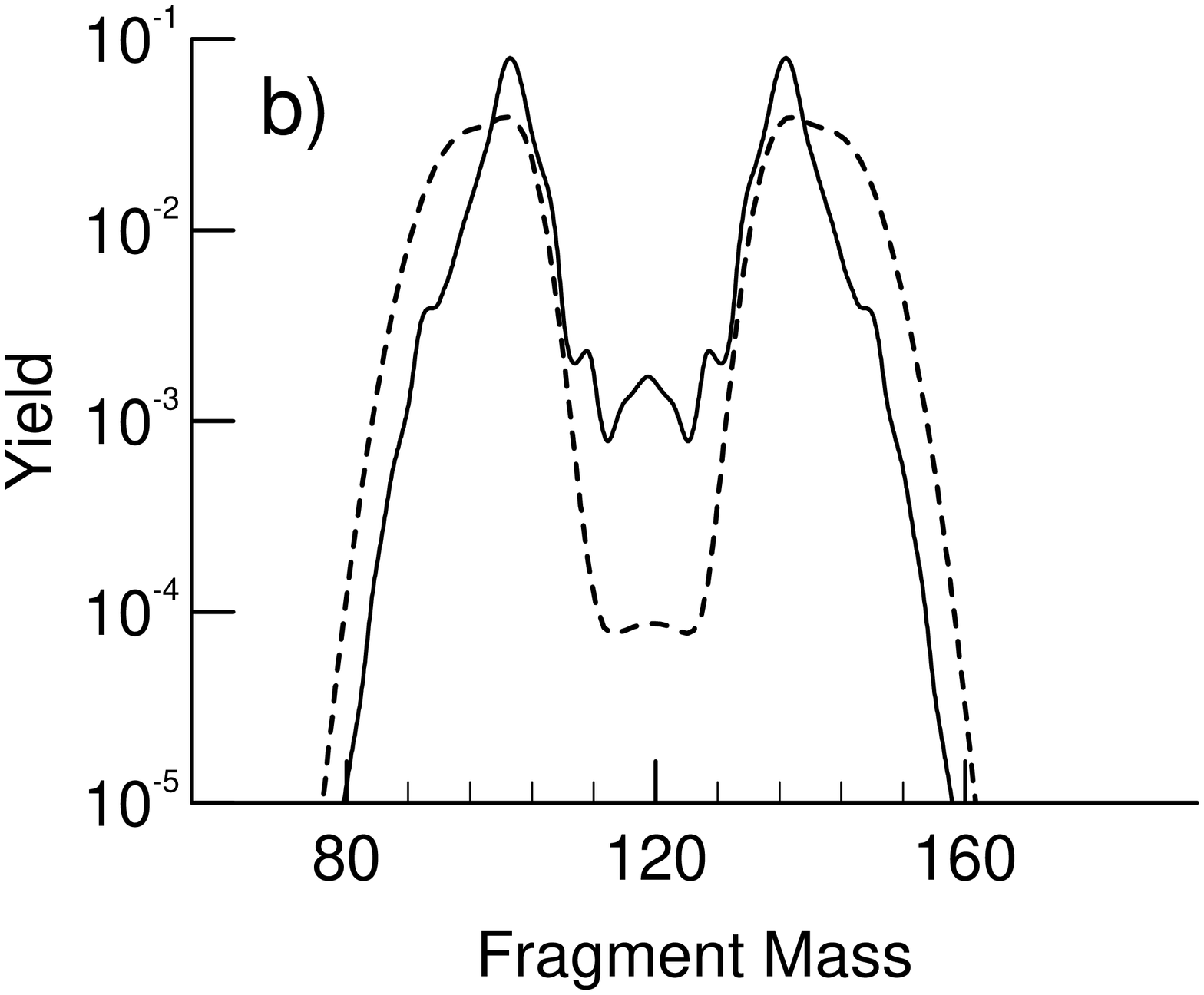}
\end{center}
\caption{Theoretical mass distributions (solid lines) are compared with the Wahl evaluations of neutron induced fission of
$^{238}$U~\cite{Wa02} (dashed lines). Excitation energies of the compound $^{238}$U nucleus measured above the barrier are
a)~E~=~2.4~MeV, b)~E~=~1.1~MeV}
\label{w1w2}
\end{figure}

One observes that the symmetric fragmentation is slightly higher at excitation energy 1.1 MeV than at 2.4 MeV.
Clearly, this approach does not reproduce an essential feature of measured or evaluated mass fragment distributions namely,
a sensitive increase of the symmetric fission yield with increasing neutron energy. As can be inferred from previous
discussions, this discrepancy is a direct consequence of the rapid decrease with increasing energy of positive parity
components in the initial state. More detailed comparisons of the theoretical predictions with Wahl evaluations~\cite{Wa02}
Fig.~\ref{w1w2} indicate, however, that the microscopic approach reproduces successfully various characteristics of the
mass distribution.

For instance, the comparison at 2.4 MeV shows that the main features of Wahl's distribution, position and height of the
maxima, ratio peak to valley and broadening of the distribution as well, are satisfactorily reproduced by the theory.
The agreement at lower energy is not as good, due essentially to the discrepancy mentioned above.

Initial conditions appear to be crucial for the prediction of mass distributions at low energy.
In view of the quality of the results presented above we consider studying more carefully this question in future works.

%%%%%%%%%%%%%%%%%%%%%%%%%%%%%%%%%%%%%%%%%%%%%%%%%%%%%%%%%%%%%%%
\subsection{Dynamical effects}

\begin{figure}[htb]
\begin{center}
\includegraphics[scale=0.3,angle=-90]{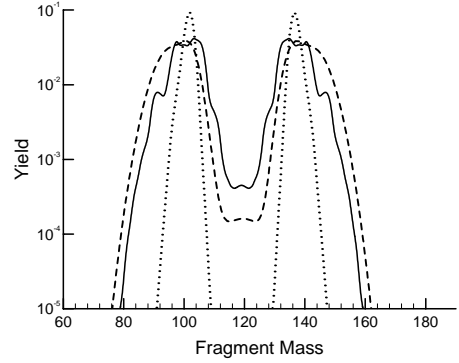}
\caption{Comparison between the "one-dimensional" mass distribution of Fig.~\ref{statique} (dotted line),
the mass distribution resulting from the dynamical calculation (solid line) with the initial state state located 2.4 MeV
above the barrier, and
the Wahl evaluation (dashed line)~\cite{Wa02}.}
\label{statedyn}
\end{center}
\end{figure}

In order to analyze the influence of dynamical effects, the fragment mass distribution obtained for the initial state
located 2.4 MeV above the first barrier shown in Fig.~\ref{w1w2}a)
is compared in Fig.~\ref{statedyn} with our previous "one-dimensional" distribution. Fig.~\ref{statedyn} also shows the
evaluated data from the Wahl systematics (dashed curve)~\cite{Wa02}.

We first note that the maxima of the two theoretical distributions are both located around $A_H$ = 134 and $A_L$ = 104,
in good agreement with evaluated data. As already mentioned in Section~\ref{static}, this is
a confirmation that the most
probable fragmentation is due essentially to shell effects in the nascent fragments and not to dynamical effects.
The widths of the peaks obtained from the full dynamical calculation are much larger- about twice as large - than those
of the "one-dimensional" one, and consequently they are in much better agreement with the Wahl evaluated data.
Clearly the dynamics is found to play a
major role in the broadening of the fragment mass distributions. In the present dynamical calculation the broadening
is clearly due to the interaction between the elongation and the asymmetry degrees of freedom, which results from
both the
potential energy and the inertia variations. As already discussed in Section~\ref{sec3}, these effects are especially
important in the descent from saddle to scission, where the inertia component $B_{23}$ is found to be large
(see Fig.~\ref{fig3}). In fact the cross term in the kinetic energy of the collective Hamiltonian appears to be
responsible
for exchanges of energy between the two modes and for the spreading of the time-dependent wave function in the
asymmetric valley.  \\
In order to quantitatively analyze those effects, the time dependent wave function $g(q_{20},q_{30},t)$
has been expanded over the "one-dimensional" states
$\Psi_n^{\pi}(q_{30})$ described in Section~\ref{static} along the line $q_{20}$ = $(q_{20})_s$ = $f(q_{30})$:
\be
\displaystyle
g((q_{20})_s,q_{30},t) =\sum_n C_n(t) \; \Psi_n^{\pi}(q_{30}) .
\ee
The weight coefficients $C_n(t)$ can be calculated as:
\be
\displaystyle
C_n(t) =\int \; dq_{30} \; g((q_{20})_s,q_{30},t) \; \Psi_n^{\pi}(q_{30}) ,
\ee
and the fraction of each "one-dimensional" state contained in the dynamical solution at scission is given by:
\be
\displaystyle
R_n(t) = \displaystyle \frac{|C_n(t)|^2}{\sum_m |C_m(t)|^2} .
\ee
Results for $R_n$(t = 0.96 $10^{-20}$ s) are listed in table~\ref{stat} in the case of the dynamical wave
function corresponding to
the initial state considered here.
\vskip 0.5 cm
\begin{table}[h]
\begin{tabular}{|l|c|c|c|c|c|} \hline
n           & 1-2   & 3-4   & 5-6   &  7-8   & 9-10  \\ \hline
$R_n$ $\%$  & 35.2    & 8.6    & 36.7    &  6.9    & 12.6     \\ \hline
\end{tabular}
\vskip 0.5 cm
\caption{Percentages of the "one-dimensional" states contained in the dynamical solution at scission for
t~=~0.96~$ 10^{-20}$~s.}
\label{stat}
\end{table}

These results indicate that
the dynamical wave function is spread over many "one-dimensional" states $\Psi_n^{\pi}$ and that
the relative contribution of the two
low-energy states is
only 35.2~$\%$. As it appears, the "one-dimensional" definition~(\ref{yield})
of the fragment yield
is not pertinent, and dynamical effects should be fully taken into
account in order to obtain realistic predictions for fragment mass
distributions.

\section{\label{sec6}Conclusion}
In this work, we have presented a theoretical framework and numerical
techniques allowing one to describe fission mass distribution in a
completely microscopic way. The method is based on a HFB description of
the internal structure of the fissioning system.
The collective dynamics is derived from a time-dependent
quantum-mechanical formalism where the wave-function of the system is
of GCM form.
A reduction of the GCM equation to a Schr\"odinger equation is made by
means of usual techniques based on the Gaussian Overlap Approximation.
Such an approach has the advantage of describing the evolution of heavy
nuclei toward fission in a completely quantum-mechanical fashion and
without phenomenological parameters. \\
Properties of the fissioning system which have a large influence on
collective dynamics have been discussed. Among them, the most important
is the variation with deformation of the nuclear superfluidity induced
by pairing correlations. In addition to strongly influence the
magnitude of the collective inertia, these correlations are essential
in our approach because they validate the adiabatic hypothesis as a
first approximation for the description of low energy fission.\\
In the present application of this method to $^{238}$U fission, two
kinds of observables have been examined and compared to experimental
data: the kinetic energy distribution and mass distribution of  fission
fragments. The kinetic energy distribution, which has been derived from
the mutual Coulomb energy of the fragments at scission, is found to be
in good agreement with data. A small discrepancy (6$\%$) is found
around the most probable fragmentation region, which could originate
from the fact that the nuclear contribution entering the mutual energy
between the two fragments is not strictly zero for the corresponding
scission configurations and that the attractive exchange Coulomb energy
between the fragments has been neglected.  Concerning fragment mass
distributions, the main result of the present study is that dynamical
effects taking place all along the evolution of the nucleus are
essential in order to obtain widths in agreement with experimental
data. In contrast, the maxima of the distributions are determined by
the static properties of the potential energy surface in the scission
region, that is by shell effects in the nascent fragments.  Finally,
the influence of the choice of the initial state has been studied. In
particular, symmetric fragment yields are found to be strongly
influenced by the parity composition of the initial state.
The quality of the results reported here encourages us to pursue
further studies of fission along these lines, with some additional
improvements. For instance, as suggested by the work cited in
Ref.~\cite{Mo04}, we cannot exclude that several valleys due to other
collective modes, such as hexadecapole or higher multipole deformation,
can appear in some fissioning systems.  Extensions to microscopic
calculations involving three or more collective coordinates are
envisaged.

\section{\label{sec7}Acknowledgment}
We would like to thank D. Bouche, N. Carjan and A. Rizea for useful advice concerning numerical methods and Professors F.
Goennenwein and K.-H. Schmidt for enlightening discussions on experimental results.
Finally, the authors wish to express their gratitude to Professor F. Dietrich and W. Younes for valuable discussions,
and for a critical review of the manuscript.
%\newpage

\appendix
\section{Hamiltonian matrix}   \label{A}

Starting from the functional Eq.~(\ref{functional}), the matrix elements of the Hamiltonian matrix K can be expressed from:
\be
\left ( \hat{H}_{coll} g \right ) (i,k,t) = \displaystyle \sum_{jl} K_{ik,jl} \; g(j,l,t) .
\label{eq0}
\ee
By writing:
\be
\displaystyle
\hat{H}_{coll}= \sum_{i,j=2}^{3}\hat{T}_{ij} + \tilde{V}(q_{20},q_{30}) \ ,
\ee
with
\be
\begin{array}{rll}
\displaystyle \hat{T}_{ij} & =&  \displaystyle -\frac{\hbar^2}{2}\frac{\partial}
{\partial q_{i0}}B_{ij}(q_{20},q_{30})
\frac{\partial}{\partial q_{j0}}   \ ,  \\
\displaystyle \tilde{V}(q_{20},q_{30})& = & \displaystyle V(q_{20},q_{30})-\sum_{i,j=2}^{3}\Delta V_{i,j}(q_{20},q_{30})  \ ,
\end{array}\label{kinetic}
\ee
\begin{widetext}
the different terms contributing to Eq.~(\ref{eq0}) are:
\begin{equation}
\begin{array}{ll}
\displaystyle (\hat{T}_{22}+\hat{T}_{33}) g(i,k,t) &=
\displaystyle \frac{1}{4\Delta q_{20}^2}
 \{ [-B_{22}(i-1,k)-B_{22}(i,k)]g(i-1,k,t)\\
&\displaystyle +[B_{22}(i-1,k)+2B_{22}(i,k)+B_{22}(i+1,k)]g(i,k,t)\\
&\displaystyle +[-B_{22}(i,k)-B_{22}(i+1,k)]g(i+1,k,t)  \} \\
&\displaystyle +\frac{1}{4\Delta q_{30}^2}
 \{[-B_{33}(i,k-1)-B_{33}(i,k)]g(i,k-1,t)\\
&\displaystyle +[B_{33}(i,k-1)+2B_{33}(i,k)+B_{33}(i,k+1)]g(i,k,t)\\

&\displaystyle +[-B_{33}(i,k)-B_{33}(i,k+1)]g(i,k+1,t)  \}   \ ,
\end{array} \label{discret1}
\end{equation}

\begin{equation}
\begin{array}{ll}
\displaystyle (\hat{T}_{23}+\hat{T}_{32}) g(i,k,t) &=
\displaystyle \frac{1}{16\Delta q_{20} \Delta q_{30}}  \{[B_{23}(i-1,k+1)+B_{23}(i,k+1)\\
&\displaystyle + B_{23}(i-1,k)+B_{23}(i,k)] g(i-1,k+1,t)\\
&\displaystyle +[B_{23}(i-1,k-1)+B_{23}(i+1,k+1)\\
&\displaystyle -B_{23}(i-1,k+1)-B_{23}(i+1,k-1)] g(i,k,t)\\
&\displaystyle +[-B_{23}(i-1,k)-B_{23}(i,k)\\
&\displaystyle -B_{23}(i-1,k-1)-B_{23}(i,k-1)]g(i-1,k-1,t) \\
&\displaystyle +[-B_{23}(i,k+1)-B_{23}(i+1,k+1)    \\
&\displaystyle -B_{23}(i,k)-B_{23}(i+1,k)] g(i+1,k+1,t) \\
&\displaystyle +[B_{23}(i,k)+B_{23}(i+1,k)\\
&\displaystyle +B_{23}(i,k-1)+B_{23}(i+1,k-1)] g(i+1,k-1,t) \}  ,
\end{array}  \label{discret2}
\end{equation}
and
\be
\tilde{V}(i,k) \; g(i,k,t)= \{V(i,k) \; -\Delta V_{22}(i,k) -\Delta V_{22}(i,k)-2\Delta V_{23}(i,k)\} g(i,k,t).
\ee
The labels $i$ and $k$ are related to $q_{20}$ and $q_{30}$, respectively, and $\Delta q_{20}$ and
$\Delta q_{30}$ are the associated discretization steps.
The following approximation has been used for the inertia term:
\be
B_{jj}(i\pm \frac{1}{2},k)  \approx  \frac{1}{2}(B_{jj}(i,k)+ B_{jj}(i\pm 1,k)) \ ,
\ee
and, for products $F$ of two functions $F_1$ and $F_2$ the following prescription has been assumed:
\be
\begin{array}{ll}
\displaystyle F(i+\frac{1}{2},k) & = \displaystyle  F_1(i+\frac{1}{2},k)F_2(i+\frac{1}{2},k)   \\
& \displaystyle \approx \frac{1}{4}(F_1(i+1,k)+F_1(i,k))(F_2(i+1,k)+F_2(i,k)) \ .
\end{array}  \label{hyp}
\ee
\end{widetext}

\end{document}